\documentclass[11pt,a4paper]{article}
\usepackage[noadjust]{cite}

\usepackage[margin=2.8cm,bottom=3.5cm]{geometry}

\usepackage{amsmath, amssymb}
\usepackage{color}
\pdfoutput=1 

\usepackage{graphicx} 
\usepackage{subcaption}
\usepackage[T1]{fontenc}

\title{Fermions on wobbling kinks: normal versus quasinormal modes}

\author{
        Jo\~ao G. F. Campos \\
        Departamento de F\'isica, Universidade Federal de Pernambuco,\\
        Av. Prof. Moraes Rego, 1235, Recife - PE - 50670-901, Brazil\\
        joao.gfcampos@ufpe.br
            \and
        Azadeh Mohammadi\\
        Departamento de F\'isica, Universidade Federal de Pernambuco,\\
        Av. Prof. Moraes Rego, 1235, Recife - PE - 50670-901, Brazil\\
        azadeh.mohammadi@ufpe.br
}

\begin{document} 
\maketitle

\begin{abstract}

The system consisting of a fermion in the background of a wobbling kink is studied in this paper.  To investigate the impact of the wobbling on the fermion-kink interaction, we employ the time-dependent perturbation theory formalism in quantum mechanics. To do so, we compute the transition probabilities between states given in terms of the Bogoliubov coefficients. We derive Fermi's golden rule for the model, which allows the transition to the continuum at a constant rate if the fermion-kink coupling constant is smaller than the wobbling frequency. Moreover, we study the system replacing the shape mode with a quasinormal mode. In this case, the transition rate to continuum decays in time due to the leakage of the mode, and the final transition probability decreases sharply for large coupling constants in a way that is analogous to Fermi's golden rule. Throughout the paper, we compare the perturbative results with numerical simulations and show that they are in good agreement.

\end{abstract}

\section{Introduction}

Solitons and solitary waves are important solutions of field theories that are topologically stable, propagate without losing the shape, and have localized energy density \cite{rajaraman1982solitons, vilenkin2000cosmic, manton2004topological}. These solutions may couple to bosons and fermions with fascinating properties such as fractional fermionic charges \cite{jackiw1976solitons,goldstone1981fractional}. In particular, kinks are solutions of scalar field theories when a discrete symmetry is broken and degenerate vacua appear. The prototypical models of kinks are the sine-Gordon, an integrable model, and the $\phi^4$ model, which is non-integrable. The system composed of a fermion interacting with background $\phi^4$ kink is analytically solvable and exhibits bound states as well as scattering states \cite{chu2008fermions, charmchi2014complete, charmchi2014massive}. The properties of fermion-kink systems, such as the fermion energy spectrum and Casimir energy, have been studied for many different models \cite{gousheh2014investigation, charmchi2014one, shahkarami2011casimir, gousheh2013casimir, bazeia2017fermionic, bazeia2019fermion}. In these systems, it is also possible to consider the back-reaction of the fermion on the soliton, which modifies the kink profile, especially for strong coupling constants \cite{shahkarami2011exact, klimashonok2019fermions, perapechka2020kinks, amado2017coupled}. 
This problem becomes even more fascinating when one considers the fermion spectrum in the presence of a static kink-antikink configuration. It was shown that the spectrum approaches two copies of the one of a single kink in the limit of large separation for both the $\phi^4$ model \cite{chu2008fermions} and the sine-Gordon \cite{brihaye2008remarks}, as expected. A more interesting problem is the time evolution of the fermion in the presence of a scalar field that is not static anymore and consists of a kink-antikink collision \cite{gibbons2007fermions, saffin2007particle, campos2020fermion}. In this case, the fermions may be transferred from one soliton to the other 
\cite{gibbons2007fermions, saffin2007particle} with decreasing probability as the system approaches the Bogomol'nyi-Prasad-Sommerfield limit, where the scalar field interaction vanishes \cite{campos2020fermion}. An exciting application to this problem appears in cosmology, where domain walls can describe branes that live in higher-dimensional universes leading to fermions localization in the extra dimensions \cite{melfo2006fermion, koley2005scalar, randjbar2000fermion}. In this picture, we could live in a domain wall \cite{rubakov1983we}, and wall collisions could cause the big-bang \cite{khoury2001ekpyrotic}. 

The kink has a shape mode in $\phi^4$ and many other scalar field models and can wobble. One of the earliest works to report this property was \cite{getmanov1976bound}, which is now well understood in terms of perturbation theory \cite{manton1997kinks, barashenkov2009wobbling, oxtoby2009resonantly}. In one of the earliest works of kink-antikink collisions, the authors considered an effective model to describe kink-antikink interactions with two collective coordinates, the position and the vibrational amplitude \cite{sugiyama1979kink}. After that, in a triplet of seminal papers \cite{campbell1986kink, peyrard1983kink, campbell1983resonance}, the authors realized that the vibrational mode existence was essential to explain the resonance phenomenon where the kink and antikink collide multiple times before separating since the translational mode exchanges energy with the vibrational mode at each bounce. The structure of the resonance was shown to have a fractal structure \cite{anninos1991fractal} and was found in several models \cite{halavanau2012resonance, simas2016suppression, demirkaya2017kink, bazeia2018scattering, gani2018scattering, lima2019boundary, bazeia2019kink, christov2021kink}. Surprisingly, the $\phi^6$, where the isolated kinks do not have a vibrational mode, also exhibits resonance windows. In this case, it was shown that the energy is stored in the vibrational mode of the compound kink-antikink configuration \cite{dorey2011kink}. Interestingly, the resonance exchange mechanism only received a satisfactorily quantitative confirmation very recently \cite{manton2021kink, manton2021collective}.

To understand the role of the shape mode at each bounce, some authors considered collisions between initially wobbling kinks and showed that the initial vibrational energy could be converted into translational energy \cite{izquierdo2021scattering, campos2021wobbling}. It is clear that the kink's vibrational mode is essential and that, even if the kink is not vibrating initially, this mode will get excited after a kink-antikink collision. Therefore, if a fermion interacts with a kink, it is important to ask how this interaction is modified when the kink starts wobbling. 

An interesting modification of the usual scalar field models occurs when the potential is constructed to turn the vibrational mode into a quasinormal mode. In fact, many models already possess a tower of QNM for the stability equations around a kink solution. The quasinormal modes of the Schr\"{o}dinger and Klein-Gordon equations can be found analytically for a few potentials, as described in \cite{boonserm2011quasi}. See also \cite{bizon2011dynamics} for a similar analysis. However, we are interested in the shape mode solution of the stability equation, which is the most relevant in the kink context. This mode can be turned into a quasinormal mode if one modifies the scalar potential, which was shown to suppress the resonance windows, as the decay rate of the quasinormal mode is increased \cite{dorey2018resonant, campos2020quasinormal}.

In this work, we consider a fermion in the presence of wobbling kinks with both normal and quasinormal modes. In our case, the wobbling kink solution is considered as a background, that is, without a back-reaction. This is a common simplifying approximation that allows analytical treatment and is almost exact, in the limit where the coupling between the fermion and the kink is small \cite{chu2008fermions, charmchi2014complete}. We will show that the transition probabilities between the fermion bound and continuum states can be computed by utilizing the time-dependent perturbation theory formalism in quantum mechanics. We find that the fermions radiate to infinity if the coupling constant is small. This effect is quantified by Fermi's Golden Rule and can be reduced if the vibrational mode is turned into a quasinormal mode. In the next section, we discuss the framework we will use to study our system. In section \ref{phi4} we apply our framework to the $\phi^4$ model and in section \ref{toy} to a toy model that exhibits quasinormal besides normal modes. Finally, in section \ref{concl}, we summarize our conclusions.

\section{Model}

Consider the following model representing a scalar field interacting with a fermion field in $(1+1)$ dimensions
\begin{equation}
\mathcal{L}=\frac{1}{2}\partial_\mu\phi\partial^\mu\phi+V(\phi)+i\bar{\psi}\gamma^\mu\partial_\mu\psi-g\phi\bar{\psi}\psi.
\end{equation}
We ignore the back-reaction of the fermion field on the scalar field, which simplifies the system considerably and makes it possible to perform some analytical analysis. This is only a good approximation to the full equations of motion when $g\ll 1$. However, we will also extrapolate our results for larger values of $g$. Besides that, it creates the possibility to explore the effect of fermion excited states, which only appear when $g\geq 1$. 

The potential is chosen such that the scalar field has a kink solution $\phi_k$. In this case, the fermion equation of motion in the presence of the kink is given by
\begin{equation}
i\gamma^\mu\partial_\mu\psi-g\phi_k\psi=0.
\end{equation}
Taking the representation $\gamma^0=\sigma_1$, $\gamma^1=i\sigma_3$ for the gamma matrices, where $\sigma_i$ are Pauli matrices and multiplying the above equation by $\gamma^0$, it becomes
\begin{equation}
\label{eom}
i\partial_t\psi=H_0\psi,
\end{equation}
with the Hamiltonian operator in the form
\begin{equation}
H_0=-i\sigma_2\partial_x+g\phi_k\sigma_1.
\end{equation}
This operator can be diagonalized by solving the Schr\"{o}dinger-like equations
\begin{equation}
-\partial_x^2\psi_{n,\pm}+V_{\pm}\psi_{n,\pm}=E_n^2\psi_{n,\pm}.
\label{fermion-spectrum}
\end{equation}
with the effective potentials given by $V_\pm=g(g\phi_k^2\mp\partial_x\phi_k)$ and the eigenfunctions by $\psi_n=(\psi_{n,+},\psi_{n,-})^T$. Because they are partner Hamiltonians \cite{cooper1995supersymmetry} it is guaranteed that both have the same eigenvalues, except for the zero mode ($E=0$), and that the eigenfunctions obey the first order equation
\begin{equation}
\label{eq_eig}
H_0\psi_n=E_n\psi_n.
\end{equation}

Now, if we perturb the kink with the shape mode, the scalar field becomes $\phi=\phi_k+A\eta_S\cos(\omega_St)$ and the Dirac equation gives
\begin{equation}
i\partial_t\psi=(H_0+AH_1(t))\psi,
\end{equation}
where the perturbation is given by $H_1(t)=g\eta_S\cos(\omega_St)\sigma_1$ and $A$ is the amplitude of the perturbation. Then, assuming that the perturbation term is small compared with $H_0$, one can write the fermion field as a linear combination of $H_0$ eigenfunctions, which in the Schr\"{o}dinger picture evolves as
\begin{equation}
\psi=\sum_nc_n(t)e^{-iE_nt}\psi_n.
\end{equation}
We expand $c_n(t)$ in powers of $A$ in the following form
\begin{equation}
c_n(t)=c^{(0)}_n(t)+Ac^{(1)}_n(t)+\cdots.
\end{equation}
Considering the initial condition $c_n(0)=\delta_{ni}$, first-order time-dependent perturbation theory gives \cite{Sakurai:1341875}
\begin{equation}
\label{eq_c1}
c^{(1)}_n(t)=-i\int_0^te^{i\omega_{ni}t^\prime}\langle n|H_1(t^\prime)|i\rangle dt^\prime,
\end{equation}
where $\omega_{ni}=E_n-E_i$ and the matrix element of any operator $\mathcal{O}$ is defined as usual
\begin{equation}
\langle n|\mathcal{O}|i\rangle=\int\psi_n^T\mathcal{O}\psi_idx.
\end{equation}
For a given $i$, the coefficients $c_n(t)$ are the Bogoliubov coefficients, which quantify the probability amplitude of the transitions between states \cite{saffin2007particle}. Thus, the transition probability for $n\neq i$ is given by
\begin{equation}
P_{i\to n}(t)=A^2|c_n^{(1)}(t)|^2.
\end{equation}
This result is valid as long as $A^2|c_n^{(1)}(t)|^2\ll 1$. We can perform the time integration in eq.~(\ref{eq_c1}) to obtain
\begin{equation}
\label{eq_cn}
c^{(1)}_n(t)=\frac{g}{2}\langle n|\eta_S\sigma_1|i\rangle\left(\frac{1-e^{i(\omega_{ni}+\omega_S)t}}{\omega_{ni}+\omega_S}+\frac{1-e^{i(\omega_{ni}-\omega_S)t}}{\omega_{ni}-\omega_S}\right).
\end{equation}
For large $t$, one can compute the transition rate from the initial to the final state. It is defined as the large $t$ limit of the ratio between the transition probability $P_{i\to n}$ and time $t$. According to Fermi's golden rule, it is given by
\begin{equation}
\label{FGR}
R_{i\to n}=A^2g^2\frac{\pi}{2}\left|\langle n|\eta_S\sigma_1|i\rangle\right|^2\left[\delta\left(\omega_{ni}+\omega_S\right)+\delta\left(\omega_{ni}-\omega_S\right)\right],
\end{equation}
for $n\neq i$, which must be integrated over a set of final states.

\section{$\phi^4$ model}
\label{phi4}

Let us start with the perturbative analysis on the fermion field interacting with the kink of the famous $\phi^4$ model where the system is analytically solvable \cite{chu2008fermions, charmchi2014complete}. We need to set $V(\phi)=\frac{1}{2}(\phi^2-1)^2$, $\phi_k=\tanh(x)$, $\eta_S=\tanh(x)\text{sech}(x)$ and $\omega_S=\sqrt{3}$. In this case, the effective fermionic potentials are in the P\"{o}schl-Teller form
\begin{equation}
\label{eq_PT}
V_\pm=g^2-\frac{g(g\pm1)}{\cosh^2(x)}.
\end{equation}
They have analytical solutions, which are listed in Appendix \ref{ap1}. In the following analysis, we will focus on the case where the initial fermion state is the zero mode with $i=0$. Using the analytical expressions for the bound and scattering states, we can compute the matrix elements $\langle n|\eta_S\sigma_1|0\rangle$.  

\begin{figure}
     \centering
     \begin{subfigure}[b]{0.48\textwidth}
         \centering
         \includegraphics[width=\textwidth]{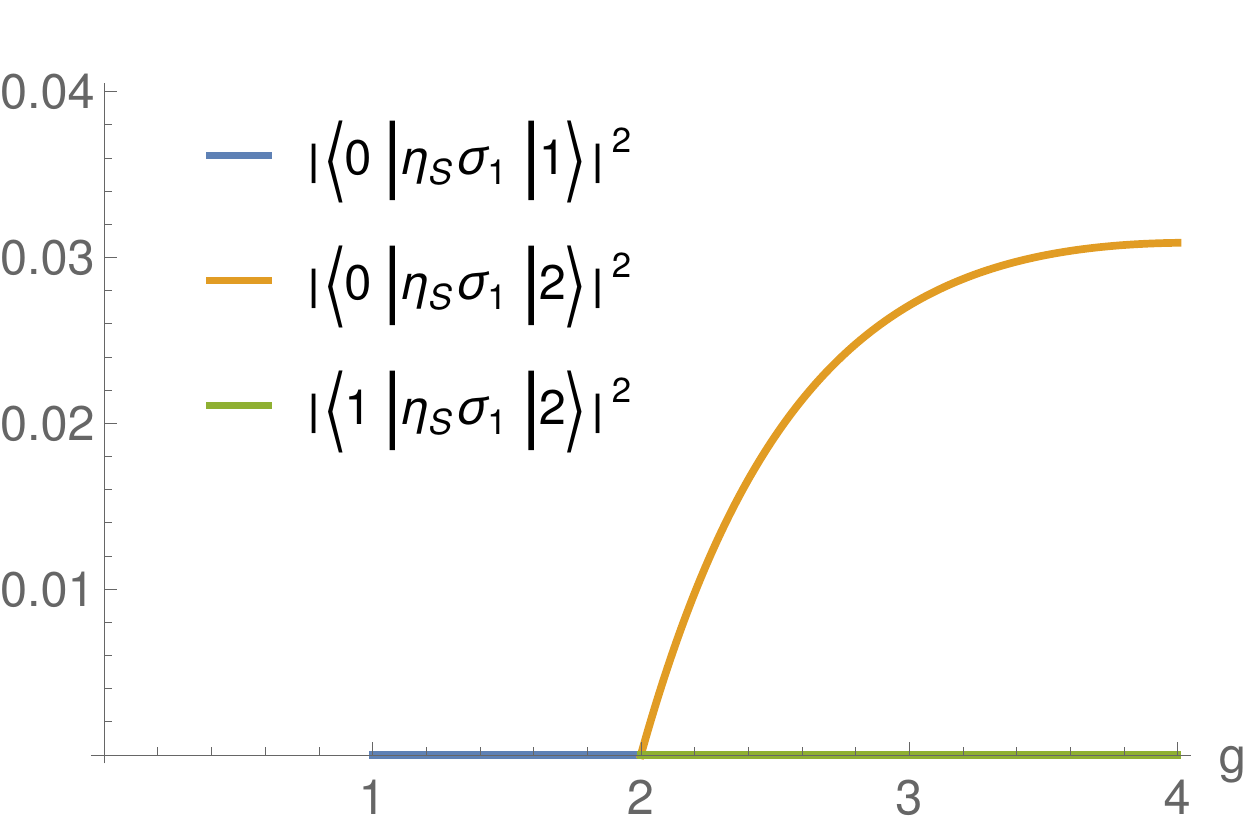}
         \caption{}
         \label{fig_etan}
     \end{subfigure}
     \hfill
     \begin{subfigure}[b]{0.48\textwidth}
         \centering
         \includegraphics[width=\textwidth]{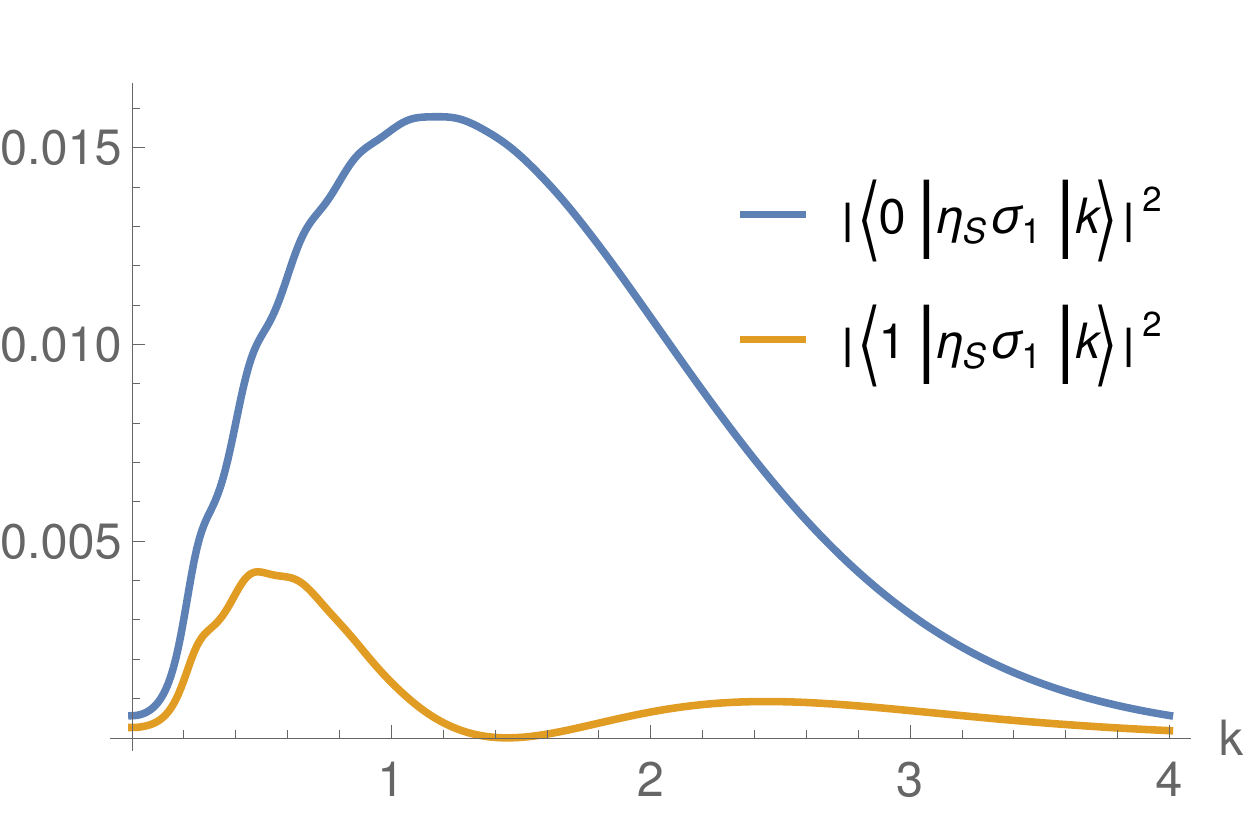}
         \caption{}
         \label{fig_etak}
     \end{subfigure}
        \caption{Matrix elements for the operator $\eta_S\sigma_1$ between the eigenstates of $H_0$ with fermion interacting with the $\phi^4$ wobbling kink. We fix $g=1.6$ in (b).}
        \label{fig_matel}
\end{figure}

For the $\phi^4$ model, a fermion bound state appears whenever $g$ is an integer. The first excited state appears when $g=1.0$, and the matrix element between this state and the zero mode is given by
\begin{equation}
\langle 1|\eta_S\sigma_1|0\rangle=\int\left(\psi_{1,-}\eta_S\psi_{0,+}+\psi_{1,+}\eta_S\psi_{0,-}\right)dx,
\end{equation}
for example. Because $\psi_{0,+}$ and $\psi_{1,-}$ are even and $\eta_S$ is odd, the first term vanishes and, because $\psi_{0,-}=0$, the second term also vanishes. In general, the integral vanishes by parity arguments whenever the difference between the index of the two bound states is an odd number. In contrast, the transition from the zero mode to the next excited state, appearing at $g=2.0$, is allowed. In Fig.~\ref{fig_matel}, some of the matrix elements $\langle n|\eta_S\sigma_1|i\rangle$ as a function of the coupling between the fermion and wobbling kink, $g$, are shown in the left panel. In the right panel, we depict the transition from the zero mode and also the first excited state to any continuum state\footnote{The continuum states are always denoted by the letter $k$.} $|k\rangle$, varying $k$. This shows the amount of fermion radiating away for the fixed value of the coupling $g=1.6$ and, in general, these transitions are allowed. We can also see that the radiation to the high continuum state, i.e., large $k$, is suppressed as the energy gap between the initial and final states becomes too large, making the transition impossible. We observe the same behavior in the other models studied here.

\begin{figure}
     \centering
     \begin{subfigure}[b]{0.46\textwidth}
         \centering
         \includegraphics[width=\textwidth]{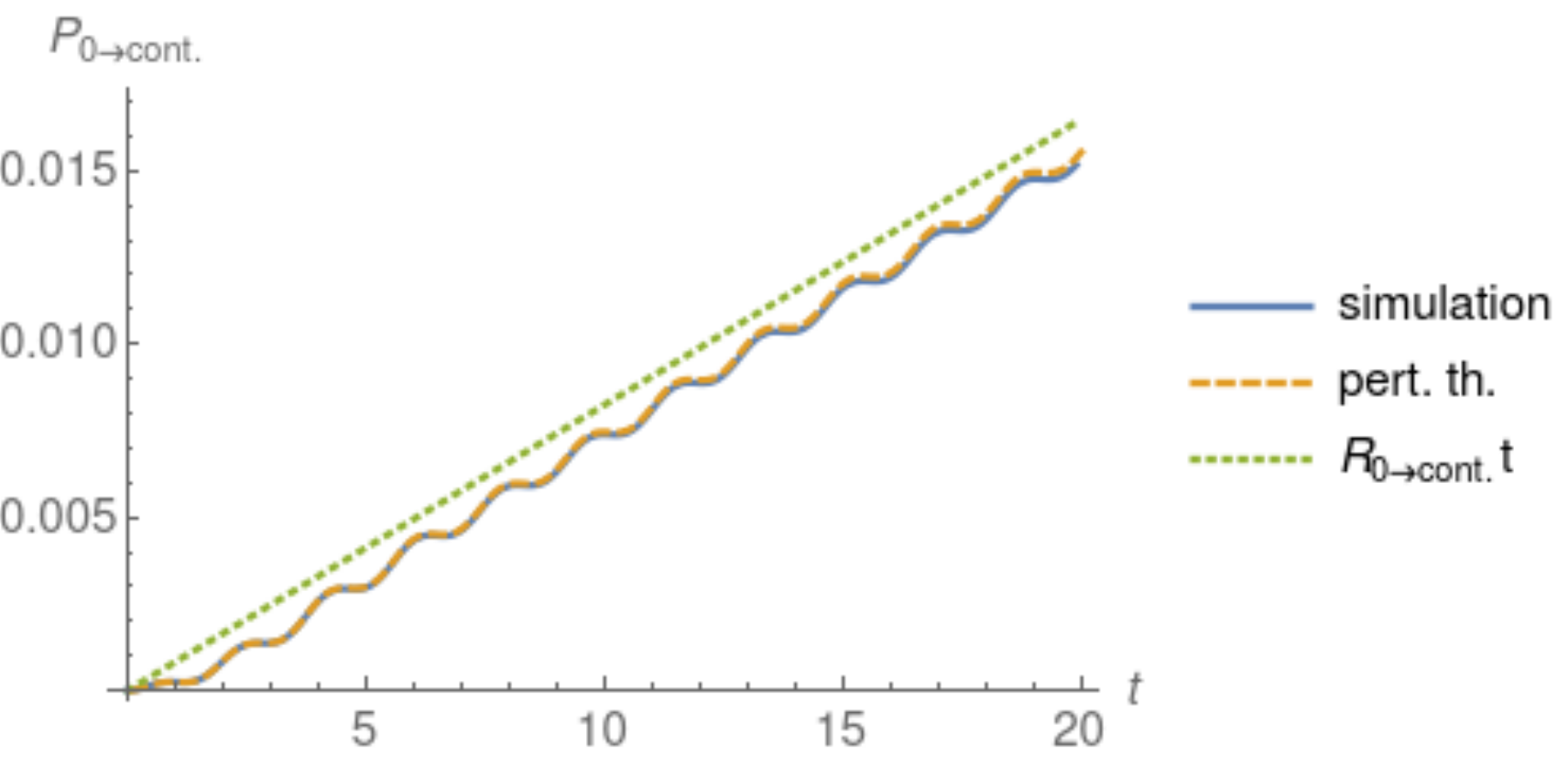}
         \caption{$g=0.8$}
         \label{fig_coef1}
     \end{subfigure}
     \hfill
     \begin{subfigure}[b]{0.50\textwidth}
         \centering
         \includegraphics[width=\textwidth]{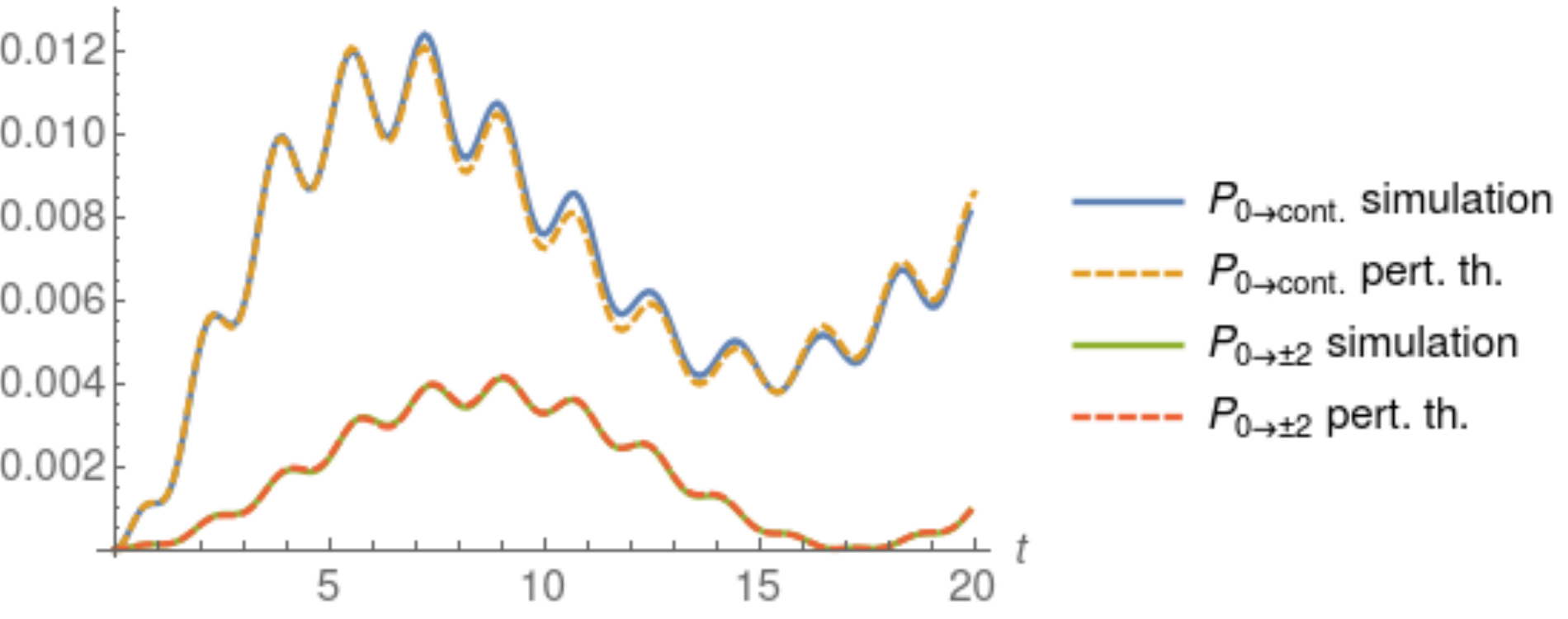}
         \caption{$g=2.1$}
         \label{fig_coef3}
     \end{subfigure}
     \caption{Transition probabilities from the zero mode to continuum and bound states for a perturbation with amplitude $A=0.1$. We compare simulation results with first-order perturbation theory. The rate from the zero mode to the continuum (eq.~(\ref{FGR2})) is also shown in (a) with the dotted line.}
        \label{fig_denscoef}
\end{figure}

The time evolution of the transition probability is shown in Fig.~\ref{fig_denscoef}. In the figure, we compare a numerical simulation of the full equations of motion (eq.~(\ref{eom})) with the approximate result of first-order perturbation theory (eq.~(\ref{eq_cn})). The numerical methods used in the simulations are described in Appendix \ref{ap2}. In these calculations we take the analytical expression for the scalar field $\phi=\phi_k+A\eta_S\cos(\omega_St)$. We set $A=0.1$ because it is of the same order of magnitude of typical wobbling amplitudes of a kink after a kink-antikink collision \cite{izquierdo2021scattering}. As one can see, the result of the simulation and the perturbative one match well.
For $g=0.8$, we compute the transition probability from the zero mode to the continuum. This is found semi-analytically by integrating eq.~(\ref{eq_cn}) over all the continuum states. We see that the probability increases at a constant rate, except for a small oscillation with frequency $\omega_S$, the kink wobbling frequency. This increase is expected for large times whenever Fermi's golden rule allows the transition to the continuum. For a final state in the continuum and the initial state in the zero mode, we have that $\omega_{ki}=\pm\sqrt{g^2+k^2}$. Thus, the transition to some state in the continuum is allowed by Fermi's golden rule whenever $g<\omega_S=\sqrt{3}$. More explicitly, we integrate eq.~(\ref{FGR}) over $k$ from $0$ to $\infty$ for the scattering states given in Appendix \ref{ap1}, $\psi_{k,L}$, with incident waves from the left. Then, including a factor of $4$ to account for the scattering states with incident waves from the right as well as the ones with negative energy, it yields
\begin{equation}
\label{FGR2}
R_{i\to \text{cont.}}=2A^2g^2\pi\bigg|\bigg\langle k=\sqrt{\omega_S^2-g^2}\bigg|\eta_S\sigma_1\bigg|i\bigg\rangle\bigg|^2\frac{\omega_S}{\sqrt{\omega_S^2-g^2}},
\end{equation}
for $g<\omega_S$ and zero otherwise. The transition rate is shown in Fig.~\ref{fig_coef1}, where one can see that it gives the correct slope of the transition probability curve. This is an interesting result because it allows one to probe whether a kink is wobbling or at least to set an upper bound on the wobbling amplitude by measuring the fermions escape rate from the kink. For $g=2.1$, we see that the transition to the continuum is not allowed anymore by Fermi's golden rule. Therefore, the transition probability to the continuum eventually stops increasing returns to zero, and this is repeated many times. Furthermore, the fermion now has a second excited state, and the transition probability to this state is finite and does not keep increasing for the same reason. In Fig.~\ref{fig_coef3}
we plot the sum of the transition probability for both second excited states with positive and negative energy. By symmetry, they both have equal transition probabilities and, therefore, the sum is just twice the transition probability to the second excited state with positive energy.

\section{Toy model}
\label{toy}

Now we repeat the same procedure for a toy model, which allows kinks with either a normal mode or a quasinormal mode \cite{campos2020quasinormal}. After rescaling, the toy model is described by the following potentials
\begin{equation} V(\phi) = 
   \begin{cases} 
      \frac{1}{2}(-\phi^2+A^2), & 0\leq\phi<\phi_1, \\
      \frac{\gamma}{2}(\phi-1)^2, & \phi>\phi_1,
   \end{cases}
\label{potentialSW}
\end{equation}
and 
\begin{equation} V(\phi) = 
   \begin{cases} 
      \frac{1}{2}(-\phi^2+A^2), & 0\leq\phi<\phi_1, \\
      \frac{\gamma}{2}[(\phi-1+\delta)^2+B^2], & \phi_1<\phi<1-\epsilon, \\
      \frac{\gamma^\prime}{2}(\phi-1)^2, & 1-\epsilon<\phi<1+\epsilon,\\
      \frac{\gamma}{2}[(\phi-1-\delta)^2+B^2], & \phi>1+\epsilon,
   \end{cases}
\label{potentialQNM}
\end{equation} 
The latter potential allows for the appearance of quasinormal modes, in contrast with the former one (\ref{potentialSW}). All quantities in the definition of the potentials are constants, except for $\phi$. The dependence on $\phi$ for negative values can be found by assuming that the potentials are even functions.

The potential (\ref{potentialSW}) is constructed such that the linearized potential of the stability equation has a square-well shape. Eq.~(\ref{potentialQNM}) modifies this potential to the square-well shape with two barriers at the sides, allowing the bound states of the previous potential to tunnel and escape from the kink if $\gamma^\prime$ is small enough. To ensure that the kinks behave correctly, we imposed continuity of the potentials and their derivatives. In this case, it is easy to show that, in the limit that $\epsilon$ goes to zero, eq.~(\ref{potentialQNM}) becomes eq.~(\ref{potentialSW}). The parameter $\epsilon$ encodes the information about the energy leak when normal mode turns into the quasinormal mode, and as it increases, the energy leakage increases. Another consequence of the continuity relations is that we have only one free parameter in eq.~(\ref{potentialSW}), which we fix by setting $\gamma=3$. This value is interesting because the kink has only a single vibrational mode and, thus, behaves similarly to the $\phi^4$ model. For eq.~(\ref{potentialQNM}), the continuity relations imply that there are three free parameters, which we fix at $\gamma=3.0$, $\gamma^\prime=1.0$ and $\epsilon=0.05$. Choosing these values, we guarantee that the only shape mode in the square-well stability potential turns into a quasinormal mode.

\begin{figure}
     \centering
     \begin{subfigure}[b]{0.48\textwidth}
       \includegraphics[width=\textwidth]{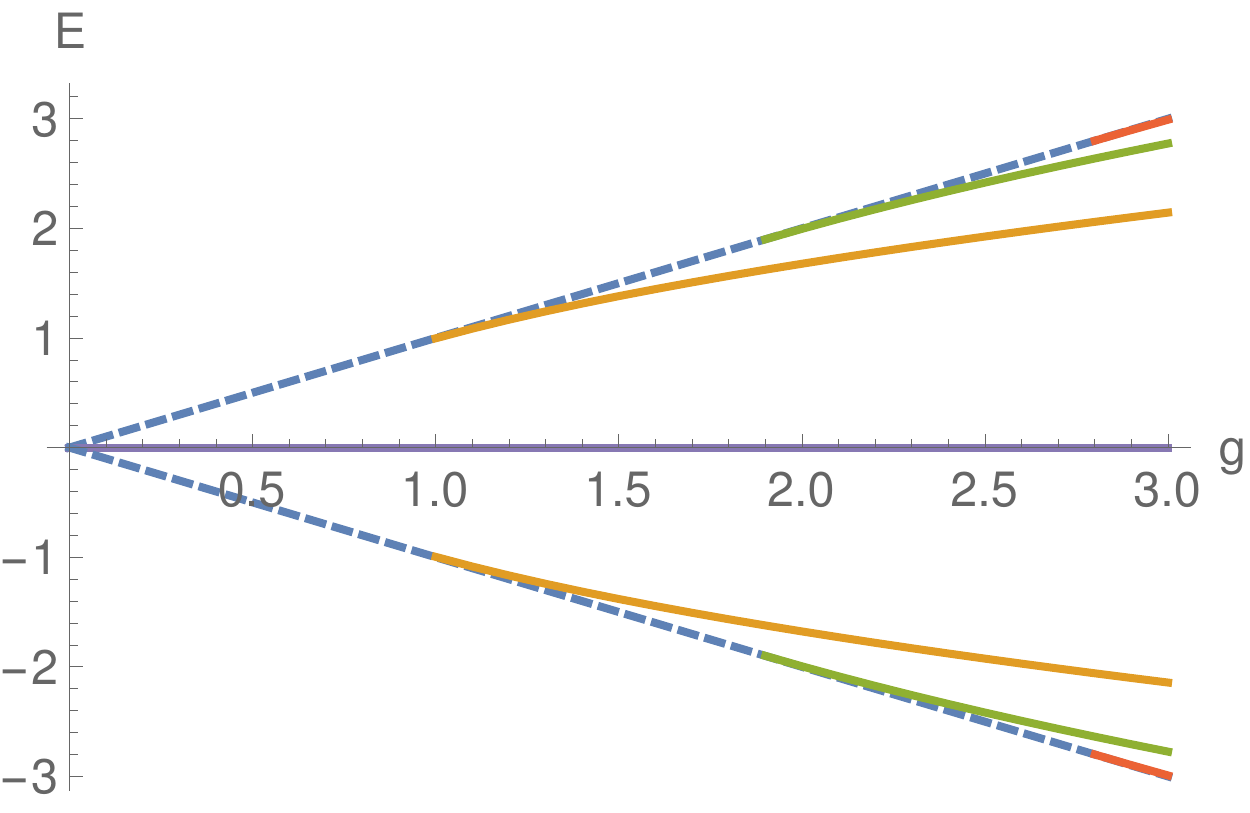}
       \caption{$\epsilon=0.0$}
       \label{fig_SWevsg}
     \end{subfigure}
     \hfill
     \begin{subfigure}[b]{0.48\textwidth}
       \includegraphics[width=\textwidth]{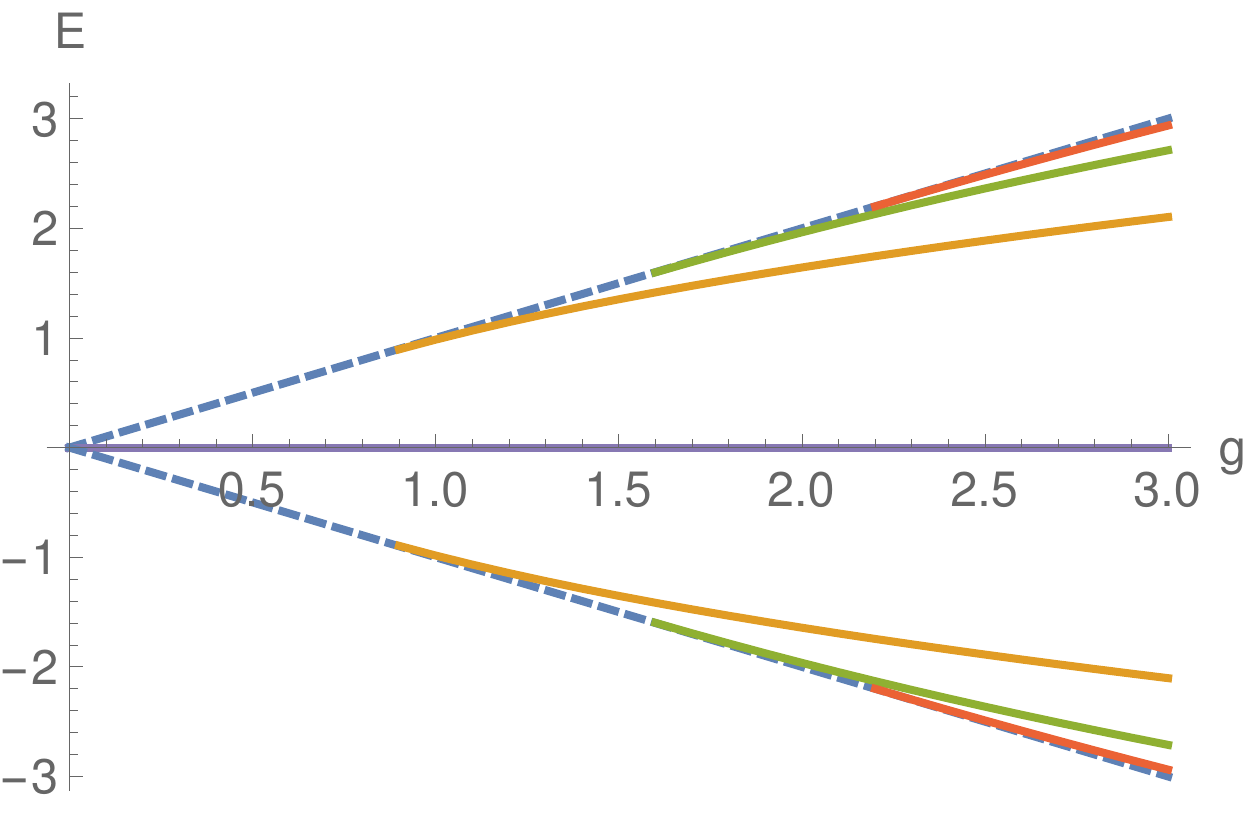}
       \caption{$\epsilon=0.05$}
       \label{fig_QNMevsg}
     \end{subfigure}
     \caption{Energy of fermion bound states as a function of coupling parameter $g$. The scalar field background is the kink of the toy model.}
     \label{fig_evsg}
\end{figure}

The kink solution for the normal mode case is given by
\begin{equation}
\phi_K(x)= 
  \begin{cases} 
     A\sin x, & 0<x<L, \\
     1-(1-\phi_1)e^{-\sqrt{\gamma}(x-L)}, & x>L,
  \end{cases}
\end{equation}
for positive $x$, where $L$ is the point where $\phi_1=A\sin(\sqrt{\lambda}L)$. For the quasinormal mode case the kink is given by
\begin{equation}
\phi_K(x)= 
  \begin{cases} 
     A\sin x, & 0<x<L, \\
     1-\delta+B\sinh(C+\sqrt{\gamma}(x-L)), & L<x<L^\prime, \\
     1-\epsilon e^{-\sqrt{\gamma^\prime}(x-L^\prime)}, & x>L^\prime,
  \end{cases}
\end{equation}
with the following definitions 
\begin{equation}
C\equiv\ln(B/D),
\end{equation}
and
\begin{equation}
D\equiv\sqrt{B^2+(1-\delta-\phi_1)^2}+1-\delta-\phi_1.
\end{equation}
Similarly, $L$ and $L^\prime$ are defined as the points where $\phi_K=\phi_1$ and $\phi_K=1-\epsilon$, respectively. The fermion spectra for both kinks, resulting from eq.~(\ref{fermion-spectrum}), are shown in Fig.~\ref{fig_evsg}. They are very similar, except that the bound states appear earlier in the quasinormal mode case. This occurs because,  in this case, the kink's tail decays more slowly, and the potentials $V_\pm$ become wider.

\begin{figure}
     \centering
     \begin{subfigure}[b]{0.48\textwidth}
         \centering
         \includegraphics[width=\textwidth]{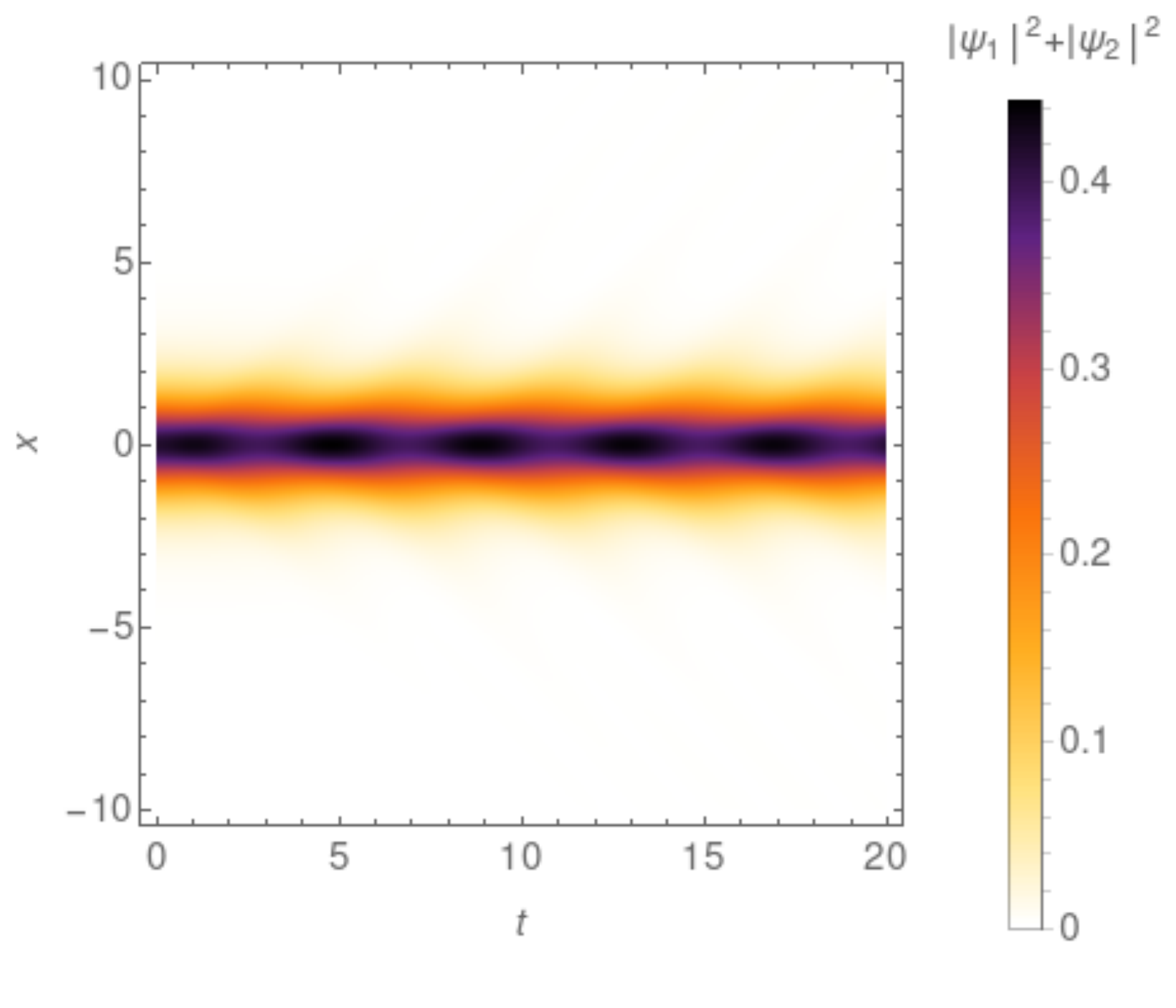}
         \caption{$\epsilon=0.0$, $g=0.8$}
         \label{fig_dens1}
     \end{subfigure}
     \hfill
     \begin{subfigure}[b]{0.48\textwidth}
         \centering
         \includegraphics[width=\textwidth]{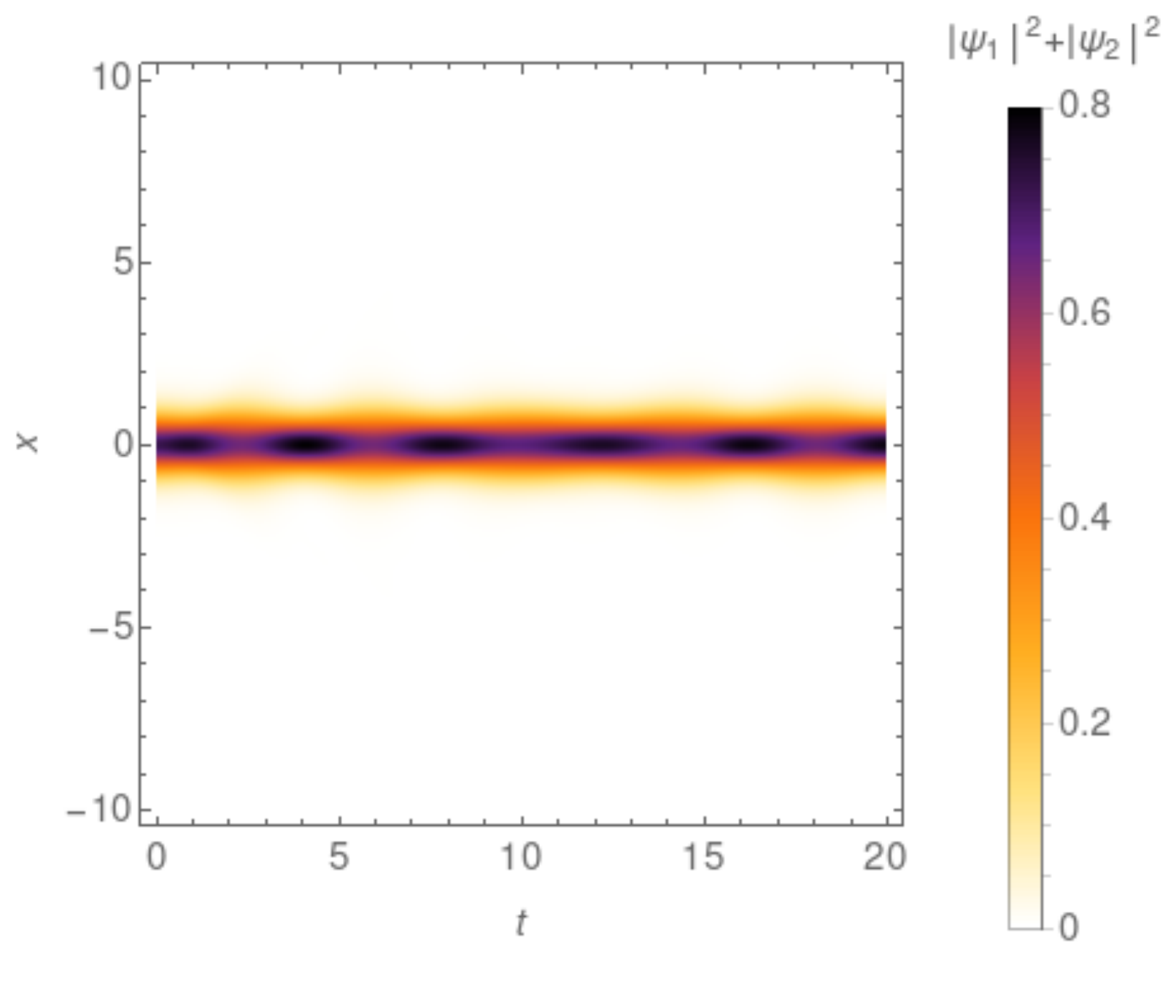}
         \caption{$\epsilon=0.0$, $g=2.1$}
         \label{fig_dens2}
     \end{subfigure}
     \hfill
     \begin{subfigure}[b]{0.48\textwidth}
         \centering
         \includegraphics[width=\textwidth]{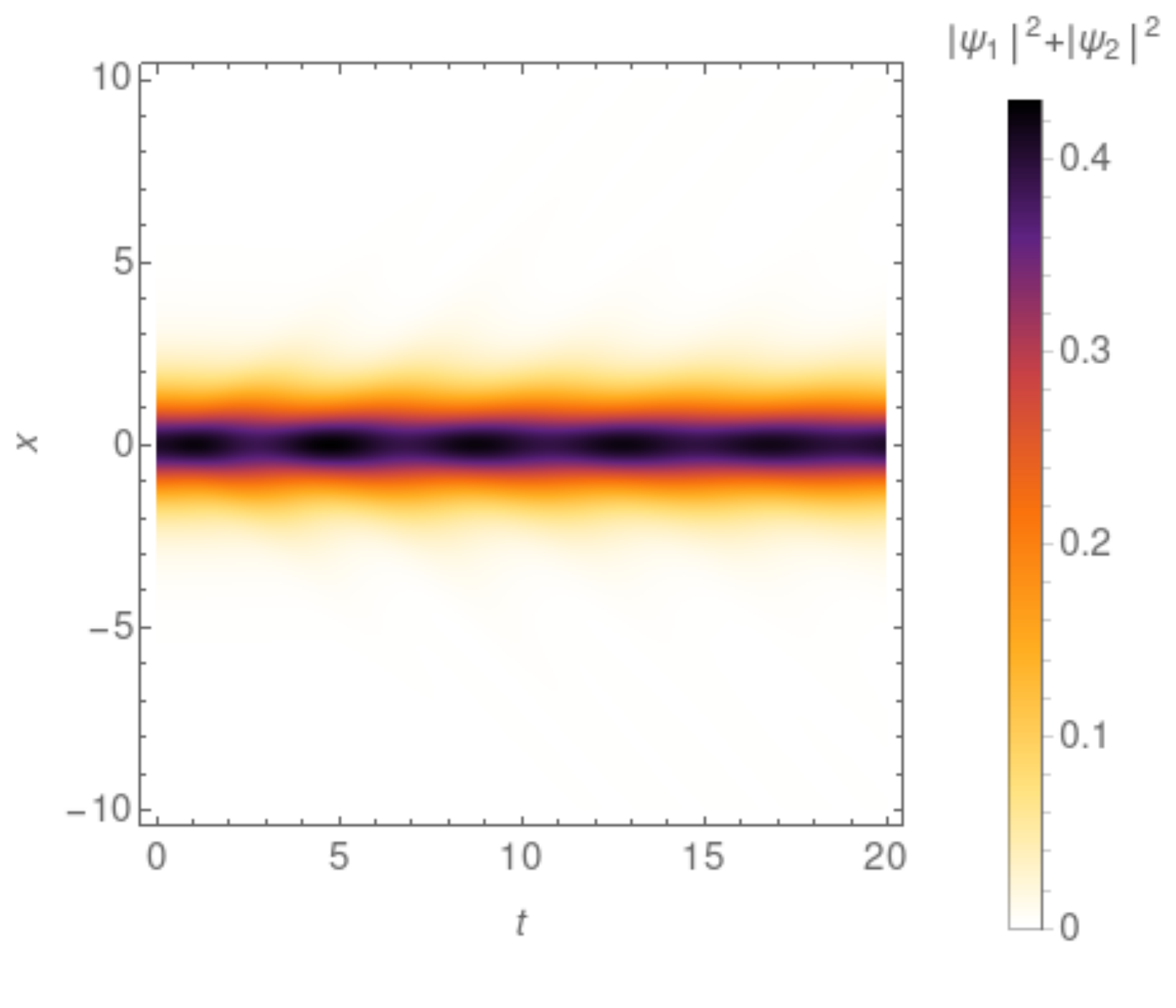}
         \caption{$\epsilon=0.05$, $g=0.8$}
         \label{fig_dens3}
     \end{subfigure}
     \hfill
     \begin{subfigure}[b]{0.48\textwidth}
         \centering
         \includegraphics[width=\textwidth]{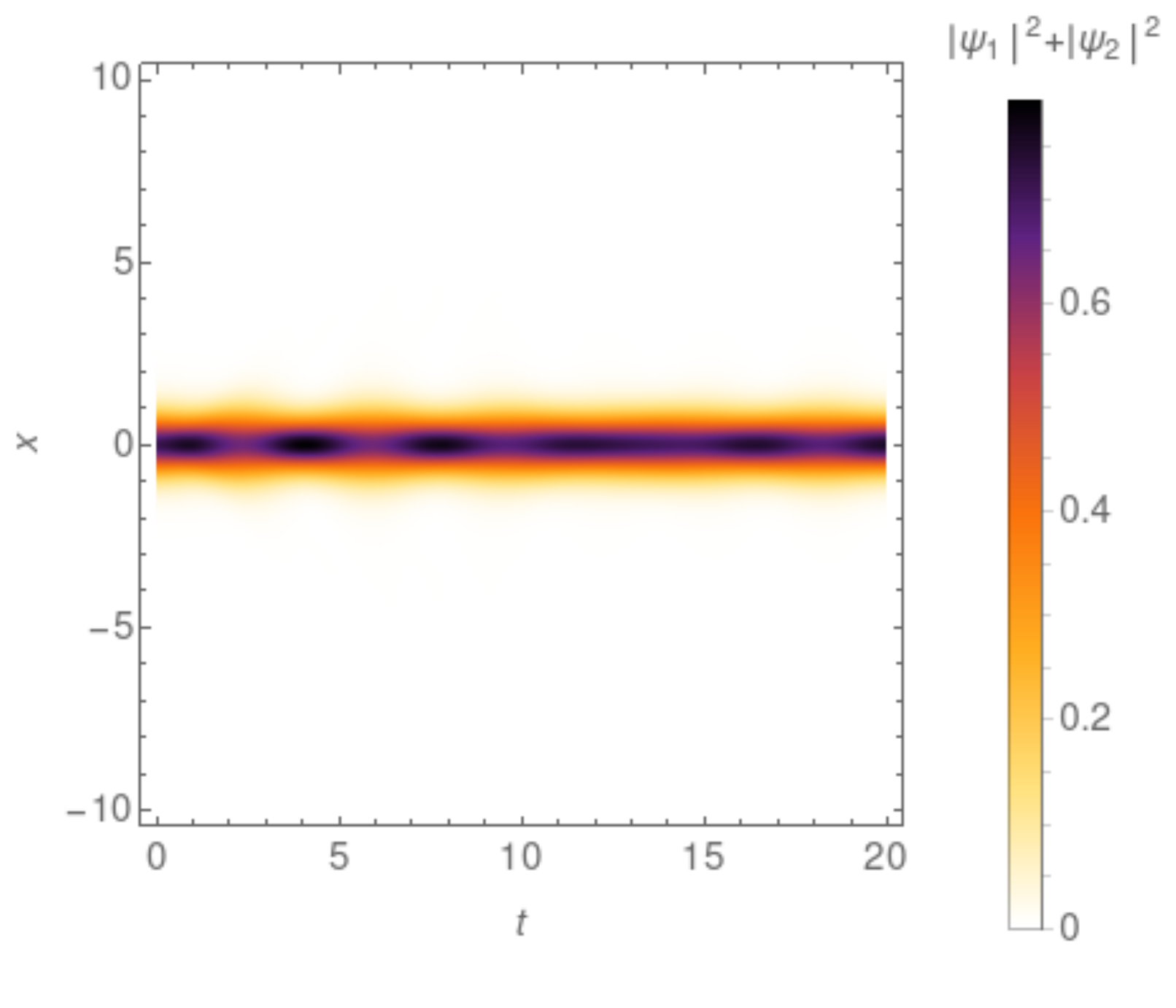}
         \caption{$\epsilon=0.05$, $g=2.1$}
         \label{fig_dens4}
     \end{subfigure}
        \caption{Evolution of the fermion density by numerically solving the equations of motion. The fermion starts at the zero mode.}
        \label{fig_dens}
\end{figure}

A typical evolution of the fermion field in the presence of a wobbling kink is shown in Fig.~\ref{fig_dens}. These were obtained by solving Dirac's equation numerically and starting the fermion at the zero mode. The plots are similar for all the models that we considered. The kink stays bound to the kink at the origin but with noticeable oscillations. As $g$ is increased, the fermion becomes more localized at the kink because the potentials $V_{\pm}$ become deeper as $g$ increases. However, the effect of the perturbations may increase with $g$ in a certain range and have the opposite effect. If the transition to the continuum is allowed according to Fermi's golden rule, it is possible to see some radiation as shown in Fig.~\ref{fig_dens1} and \ref{fig_dens3}. For the quasinormal case, the oscillations of the fermion field decrease slowly due to a decrease in the amplitude of quasinormal mode perturbation. This occurs because the quasinormal mode can tunnel the potential barrier created by the kink and radiate to infinity. However, the decay is slow and hence difficult to observe in Fig.~\ref{fig_dens3} and \ref{fig_dens4}.

We will discuss the perturbed system for both normal and quasinormal mode cases in detail and compare both numerical and perturbative solutions in the following subsections.

\subsection{Normal mode}

\begin{figure}
     \centering
     \begin{subfigure}[b]{0.48\textwidth}
         \centering
         \includegraphics[width=\textwidth]{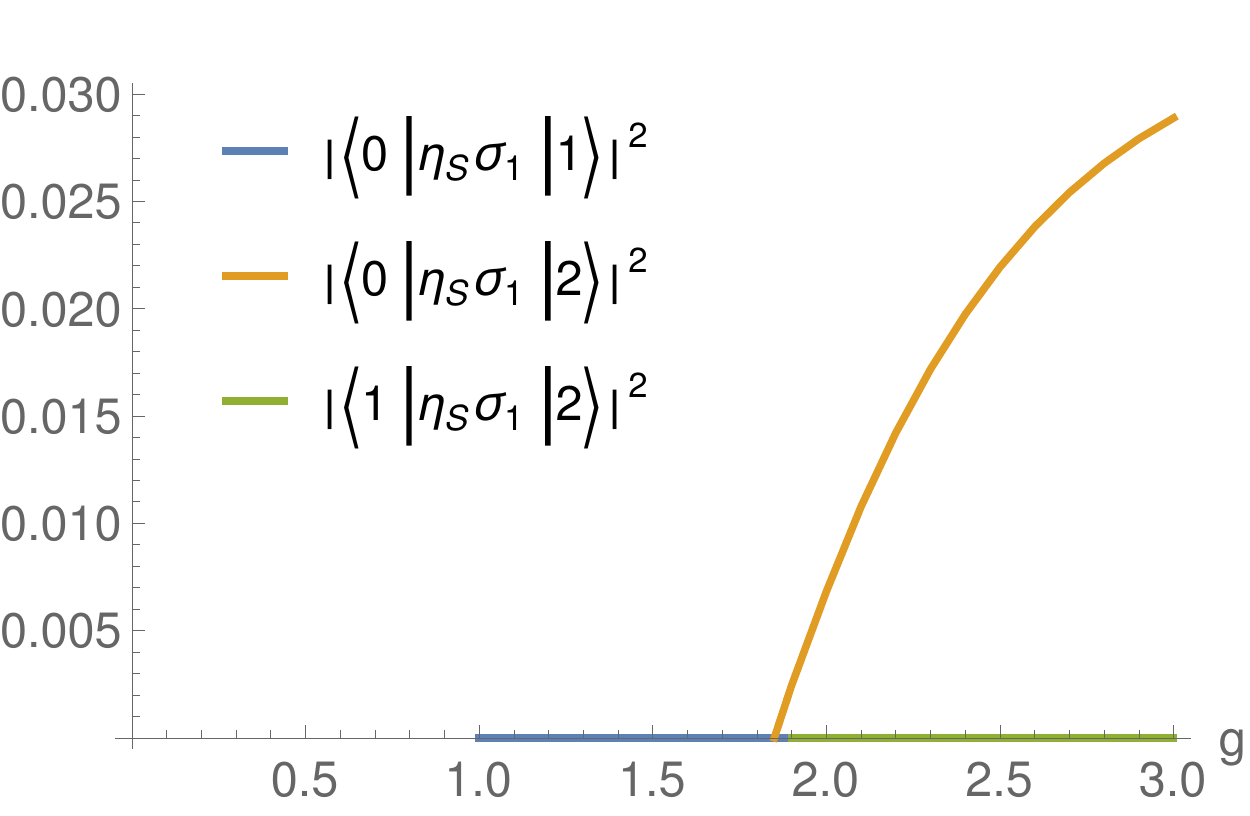}
         \caption{}
         \label{fig_etan2}
     \end{subfigure}
     \hfill
     \begin{subfigure}[b]{0.48\textwidth}
         \centering
         \includegraphics[width=\textwidth]{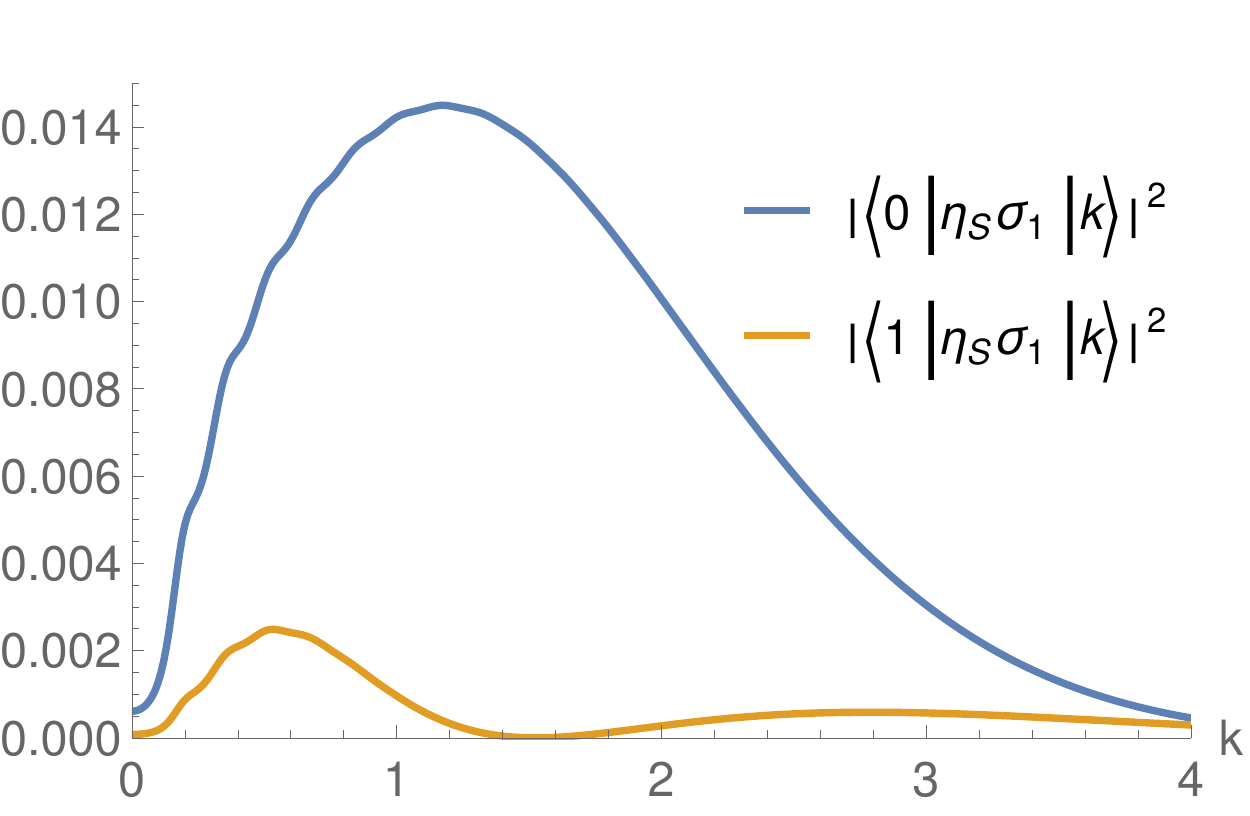}
         \caption{}
         \label{fig_etak2}
     \end{subfigure}
        \caption{Matrix elements for the operator $\eta_S\sigma_1$ between the eigenstates of $H_0$ for the toy model. The parameters are $\gamma=3.0$ and $\epsilon=0.0$. We fix $g=1.6$ in (b).}
        \label{fig_matel2}
\end{figure}

As we mentioned before, considering $\gamma=3.0$, the system possesses one shape mode, which can be found analytically \cite{campos2020quasinormal}. It has the following form with arbitrary normalization
\begin{equation}
\label{eigenodd}
\eta_S(x)= 
  \begin{cases} 
     -e^{k_Sx}, & x<-L, \\
     e^{-k_SL}\sin(p_Sx)/\sin(p_SL), & -L<x<L, \\
     e^{-k_Sx}, & x>L,
  \end{cases}
\end{equation}
where we defined $k_S\equiv\sqrt{\gamma-\omega_S^2}$ and $p_S\equiv\sqrt{\omega_S^2+1}$ with $\omega_S$ being the solution of the transcedental equation $k_S=-p_S\cot(p_SL)$. For $\gamma=3.0$ we find $\omega_S\simeq1.568$. The matrix elements of the operator $\eta_S\sigma_1$ between the eigenstates of $H_0$ are shown in Fig.~\ref{fig_matel2}. It is clear that the result is similar to the $\phi^4$ case.

\begin{figure}
     \centering
     \begin{subfigure}[b]{0.46\textwidth}
         \centering
         \includegraphics[width=\textwidth]{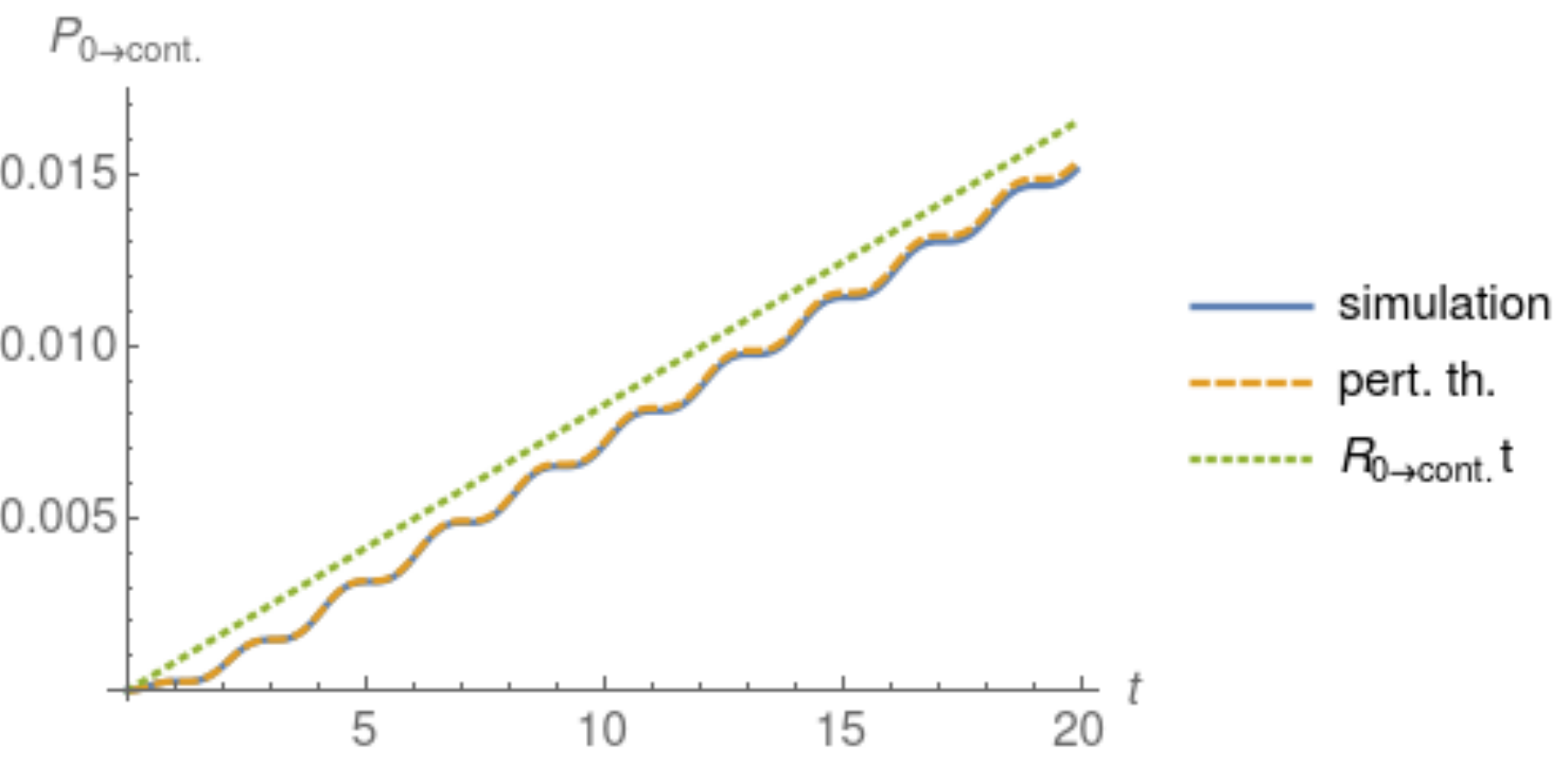}
         \caption{$g=0.8$}
         \label{fig_coef4}
     \end{subfigure}
     \hfill
     \begin{subfigure}[b]{0.50\textwidth}
         \centering
         \includegraphics[width=\textwidth]{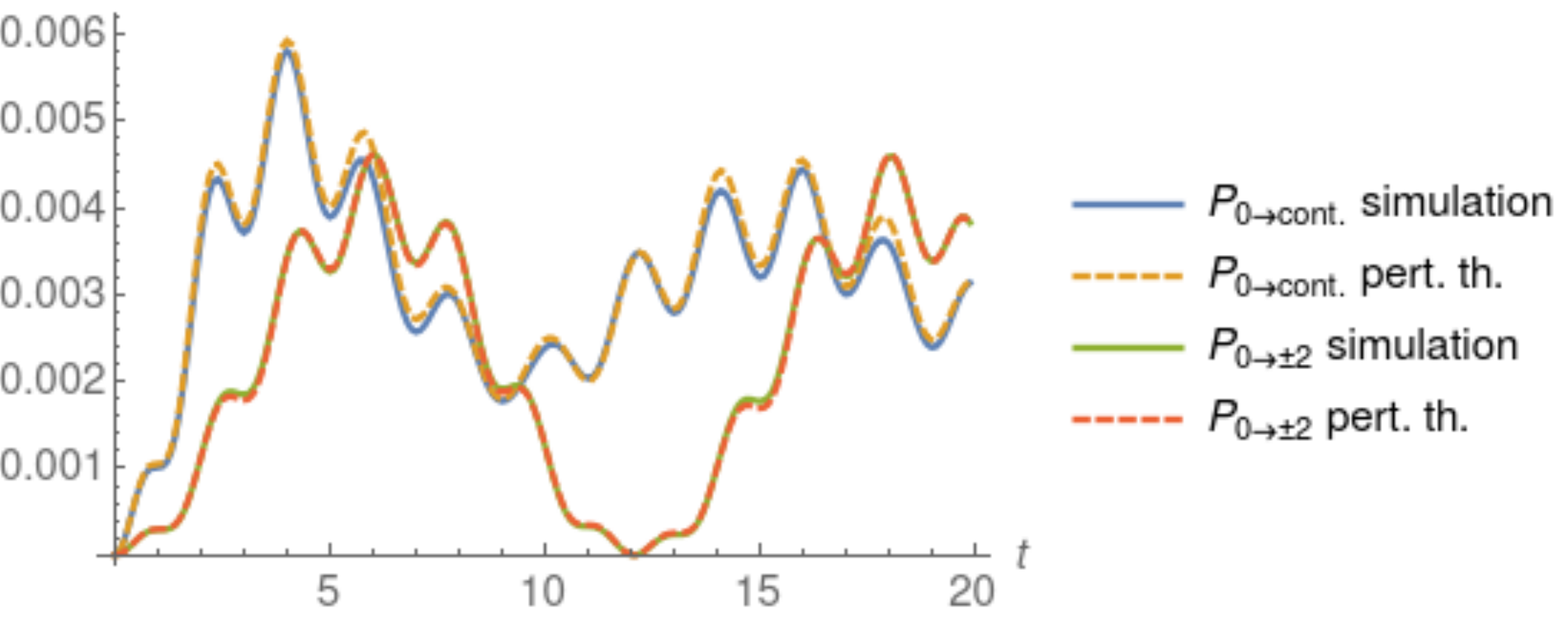}
         \caption{$g=2.1$}
         \label{fig_coef6}
     \end{subfigure}
        \caption{Transition probabilities from the zero mode to continuum and bound states for the toy model. We consider the normal mode case with parameters $A=0.1$, $\gamma=3.0$. We compare simulation results with first order perturbation theory. The rate from the zero mode to the continuum (eq.~(\ref{FGR2})) is also shown in (a) with dotted line.}
        \label{fig_denscoef2}
\end{figure}

Starting with the fermion on the zero mode we can evolve the system integrating the equations of motion subject to the perturbation numerically. These results are shown in Fig.~\ref{fig_denscoef2}, where we compare them with the analytical expressions. Again we find results very similar to the $\phi^4$ model, including the increase of the transition probability for $g<\omega_S$ with transition rate given by the Fermi's Golden Rule expression in eq.~(\ref{FGR2}). In fact, we expect similar results for a large class of models with kink solutions containing a shape mode.

The main difference between the $\phi^4$ and the toy models is that $H_0$ has to be diagonalized numerically in the latter. We find the bound states using the NDEigensystem method in Mathematica and the solve\_bvp method of the Python library SciPy. The continuum states are found as described in \cite{gousheh2013casimir}. The method deals with the scattering states instead of the parity eigenstates. However, a simple linear combination of the scattering states gives the parity eigenstates. We consider, for instance, the scattering state with an incoming wave from the left with wavenumber $k$ and energy $E_k=\sqrt{g^2+k^2}$. In short, the method consists of factoring out the highly oscillatory transmitted wave dependence as follows
\begin{equation}
\psi_k=e^{ikx}\chi_k(x).
\end{equation}
Then, we find the initial condition for $\chi_k$ by taking the limit $x\to\infty$ in the eigenvalue equation (\ref{eq_eig}). Using $\phi_k(\infty)=1.0$ and $\chi_k^\prime(\infty)=0$, this results in
\begin{equation}
\chi_k(\infty)=\begin{pmatrix}1\\
\frac{g+ik}{\sqrt{g^2+k^2}}
\end{pmatrix}.
\end{equation}
Numerically, we set infinity at $x=20.0$, and we integrate $\chi_k$ up to $x=-20.0$ according to the eigenvalue equation (\ref{eq_eig}). After that, we multiply all eigenfunctions by appropriate normalization constants.

\subsection{Quasinormal mode}

For the toy model with potential (\ref{potentialSW}), there is already a tower of quasinormal modes of the stability equation. This is also true for the $\phi^4$ model with non-integer $g$. However, physically, the most important solution of the stability equation is the shape mode for the kink systems. We want this mode to be turned into a quasinormal mode, and this can be achieved by considering potential (\ref{potentialQNM}). In this case, it is possible to find this mode analytically, and it could be used as our perturbation to the scalar field if carefully truncated. This method can work well for large times. However, as we are interested in the whole transition process and not only the large-time behavior, a more sensible approach is to use the shape mode of the model with $\epsilon=0.0$ as the initial condition of the perturbation for the $\epsilon\neq 0$ case. In this work, we will adopt this approach. The kink's shape mode becomes a quasinormal mode because the linearized potential of the normal mode case has been modified such that this mode can tunnel through the potential barrier. Therefore, this perturbation describes precisely the quasi-bound state localized at the kink, which decays in time according to the decay rate of the quasinormal mode. 

Writing $\phi=\phi_k+A\eta(x,t)$, the evolution of this perturbation can then be found numerically using the time-dependent version of the stability equation 
\begin{equation}
\label{eq_time_stab}
\frac{\partial^2\eta(x,t)}{\partial t^2}-\frac{\partial^2\eta(x,t)}{\partial x^2}+U(x)\eta(x,t)=0,
\end{equation}
where $U(x)$ is the linearized potential. In this case the perturbed Hamiltonian becomes $H_1(t)=g\eta(x,t)\sigma_1$. If we neglect the radiation generated by the evolution, we can approximate the perturbation by $\eta(x,t)\simeq\eta_S\cos(\Omega t)e^{-\Gamma t}$, where $\eta_S$ is given by eq.~(\ref{eigenodd}) and $\Omega$ and $\Gamma$ are the analytical values of the frequency and decay rate of the quasinormal mode. For $\gamma=3.0$, $\gamma^\prime=1.0$ and $\epsilon=0.05$, we find $\Gamma\simeq0.0564$ and $\Omega\simeq1.584$ according to the analytical expression in \cite{campos2020quasinormal}. In this case, eq.~(\ref{eq_cn}) is modified to
\begin{equation}
\label{eq_QNMcn}
c^{(1)}_n(t)=\frac{g}{2}\langle n|\eta_S\sigma_1|i\rangle\left(\frac{1-e^{i(\omega_{ni}+\Omega+i\Gamma)t}}{\omega_{ni}+\Omega+i\Gamma}+\frac{1-e^{i(\omega_{ni}-\Omega+i\Gamma)t}}{\omega_{ni}-\Omega+i\Gamma}\right).
\end{equation}
If we wish to include the effect of the radiation, we should take eq.~(\ref{eq_c1}) to compute $c^{(1)}_n$ with the full numerical solution of the scalar field perturbation as described above.

\begin{figure}
     \centering
     \begin{subfigure}[b]{0.46\textwidth}
         \centering
         \includegraphics[width=\textwidth]{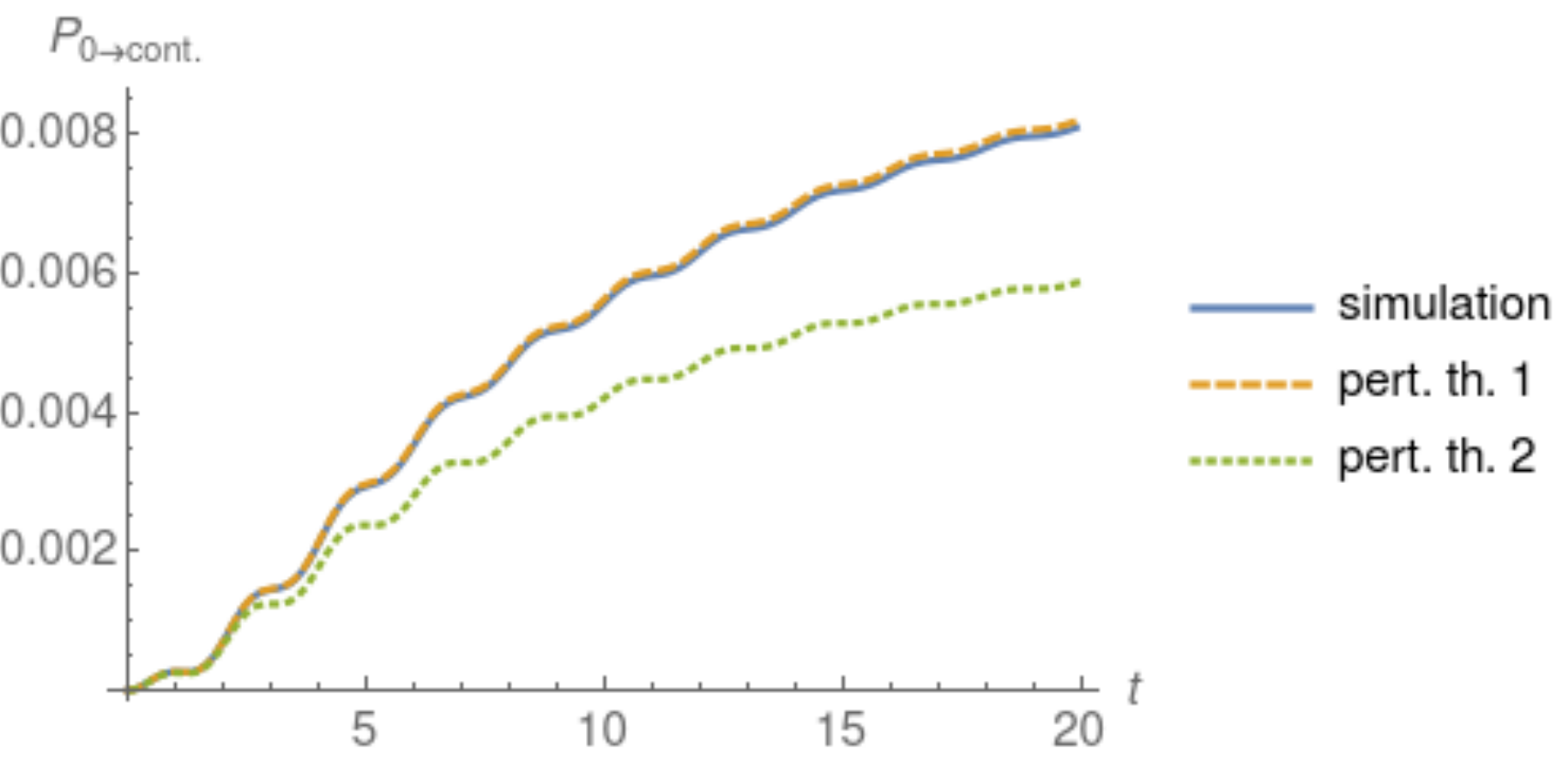}
         \caption{$g=0.8$}
         \label{fig_coef7}
     \end{subfigure}
     \hfill
     \begin{subfigure}[b]{0.50\textwidth}
         \centering
         \includegraphics[width=\textwidth]{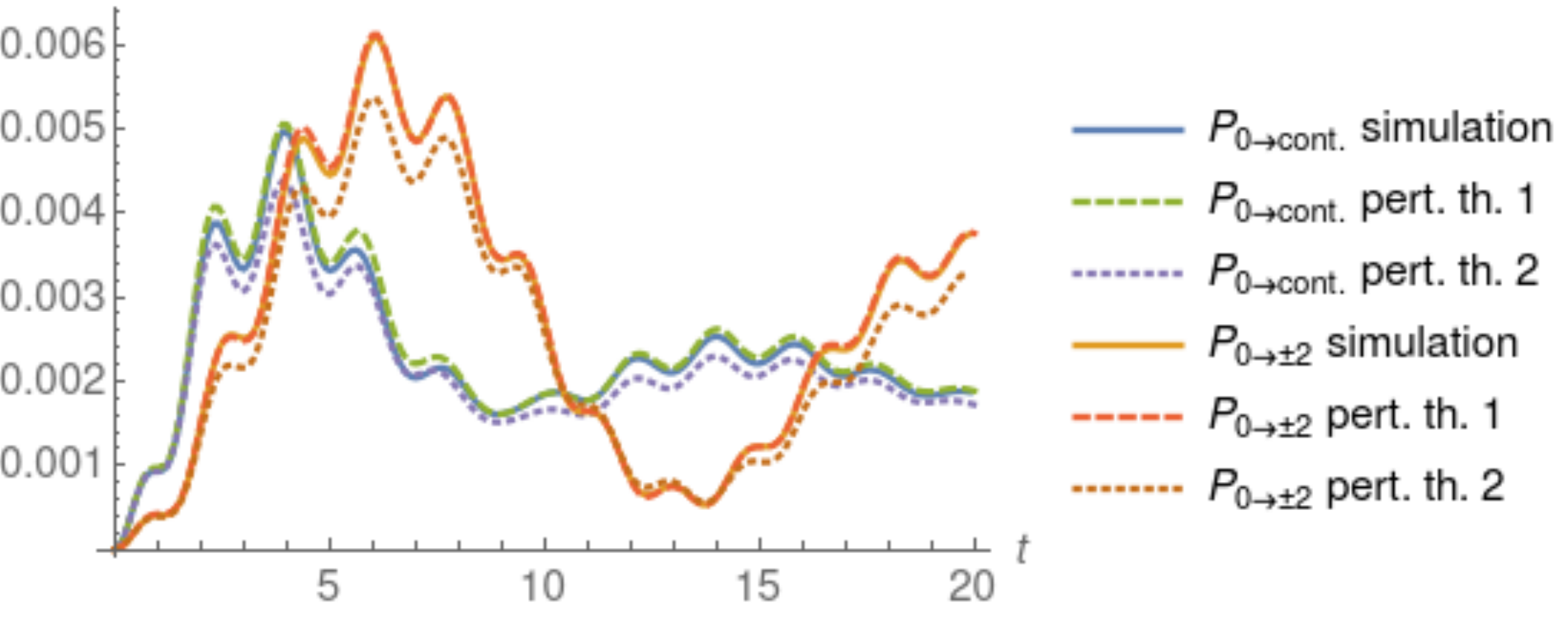}
         \caption{$g=2.1$}
         \label{fig_coef9}
     \end{subfigure}
        \caption{Transition probabilities from the zero mode to continuum and bound states for the toy model. We consider the quasinormal mode case with parameters $A=0.1$, $\gamma=3.0$, $\gamma^\prime=1.0$ and $\epsilon=0.05$. We compare simulation results with first order perturbation theory. The perturbative results are given by eq.~(\ref{eq_c1}) with the full numerical evolution of the scalar field (dashed) and by eq.~(\ref{eq_QNMcn}) (dotted).}
        \label{fig_coefs3}
\end{figure}

In Fig.~\ref{fig_coefs3} we show the time evolution of the transition probability. It behaves similarly to the previous cases, but the effect of the perturbation decays gradually in time due to the leak, which is expected from eq.~(\ref{eq_QNMcn}). In the figure, we compare the numerical simulations with both perturbative results using eq.~\ref{eq_QNMcn} and eq.~(\ref{eq_c1}) with the inclusion of radiation. The former does not have a good agreement for small values of $g$, where the effect of the radiation is more relevant but improves as $g$ is increased.

\begin{figure}
     \centering
     \includegraphics[width=0.6\textwidth]{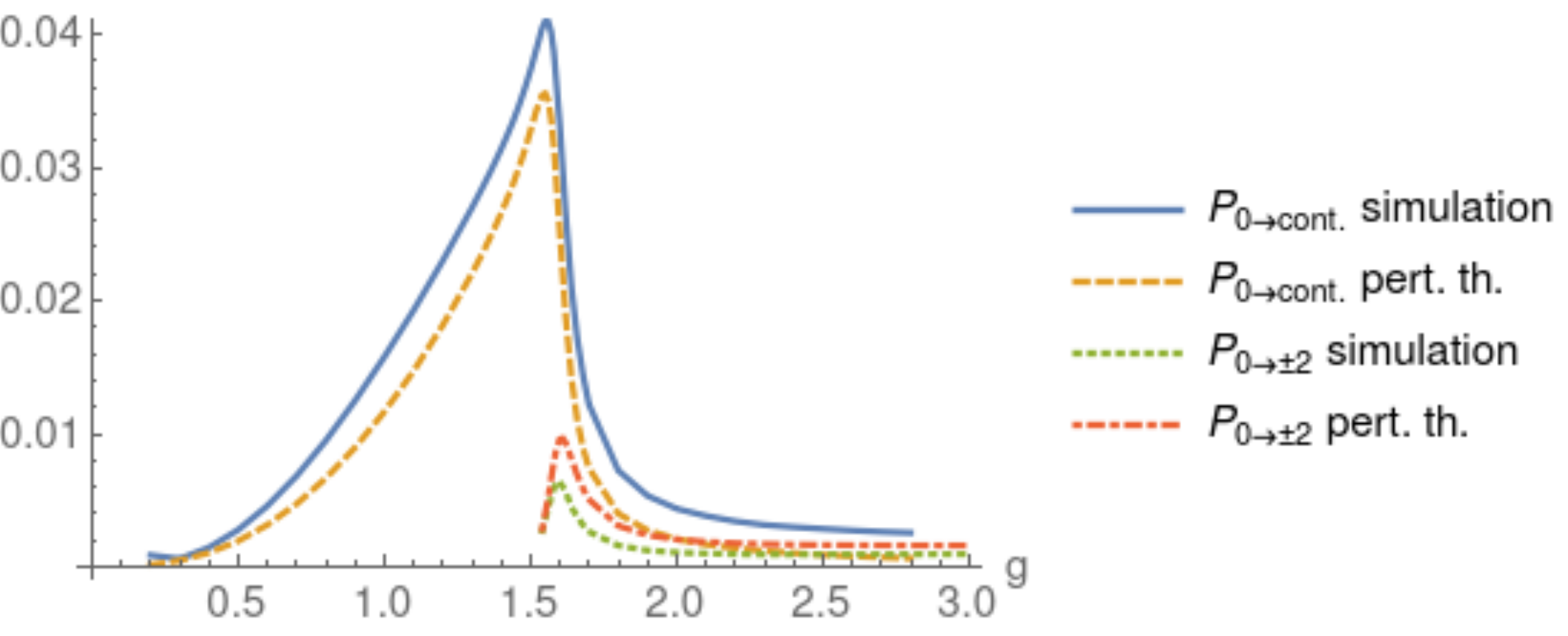}
     \caption{Asymptotic value of the transition probabilities as a function of $g$. We compare simulation results with first order perturbation theory. We consider the quasinormal mode case with parameters $A=0.1$, $\gamma=3.0$, $\gamma^\prime=1.0$ and $\epsilon=0.05$.}
     \label{fig_FS}
\end{figure}

Due to the decrease of the quasinormal mode's amplitude, the transition probabilities will reach an asymptotic value if we wait long enough. This can be used to make a more detailed analysis of the system's dependence on the coupling parameter $g$. The result is shown in Fig.~\ref{fig_FS}, where we plot the asymptotic value of the transition probability as a function of $g$. The transition probability from the zero mode to the continuum, $P_{0\to \text{cont.}}$, initially increases with $g$ due to the dependence of $H_1$ on this parameter. After a critical value of $g$, the transition becomes forbidden because the energy gap between the zero mode and the continuum becomes too large, and consequently, the transition probability decreases considerably. This occurs for $g\simeq\Omega$ and is an effect analogous to Fermi's golden rule for normal modes. Near $g=1.54$, the second excited state appears, and there is a finite transition probability to this state, which has a small peak and then also decreases considerably as $g$ increases due to the large energy gap between the states.

\section{Conclusion}
\label{concl}

In this paper, we considered a system composed of a fermion in the presence of a wobbling kink and treated the wobbling as a perturbation over the background kink configuration. We were able to write the Dirac equation for the fermion field interacting with the static kink background in the Hamiltonian formalism and diagonalize the unperturbed Hamiltonian. Moreover, we studied the fermion interaction with the wobbling $\phi^4$ kink and also kinks of a toy model with excited normal and quasinormal modes, which have analytical expressions. The toy model potential possesses a parameter $\epsilon$, which encodes the transition from the normal mode of the kink to a quasinormal mode as it becomes nonzero. As $\epsilon$ approaches zero, the toy model behaves similarly to the $\phi^4$ model \cite{campos2020quasinormal}. For the $\phi^4$ model, we used the analytical results for the energy and eigenstates, while for the toy model, we diagonalized the Hamiltonian numerically. In both cases, the threshold energy increases linearly with $g$, and more bound states with positive and negative energy appear when $g$ is increased. After diagonalizing, we include the perturbation and use the first-order perturbation theory to find the transition probabilities.  

The perturbation theory results depend on the matrix elements of the Hamiltonian perturbation, which can be easily computed in all cases considered. The matrix elements from the bound states to the continuum are suppressed when the energy gap becomes too large. Moreover, we found that these matrix elements obey parity selection rules. The time dependence of the transition probabilities can be computed analytically, except for the quasinormal mode case, where the evolution is computed numerically. From the first-order perturbation theory results, we derived Fermi's golden rule for our system, which appears because the perturbation is periodic. It states that the probability rate to the continuum vanishes if the fermion-kink coupling $g$ is greater than the wobbling frequency. The transition to the continuum was obtained by integrating the transition probability and transition rate over all the final momenta $k$ for both positive and negative energy. For the transition rate, the integral was performed explicitly.

The perturbative results were compared with numerical simulations of the Dirac equation, which showed a satisfactory agreement. In the models containing a normal mode, the transition probability to the continuum shows a steady increase when $g<\omega_S$, in agreement with Fermi's golden rule. For larger $g$, the transition probabilities oscillate around an average value, leading to a vanishing transition rate over a large time interval. The simulations show the evolution of the fermion density with time. The fermion is mostly bound to the kink and, ignoring the effect of perturbation, it becomes more localized around the kink center as $g$ is increased. On the other hand, the effect of the perturbation can be opposite to that in a certain range of $g$. Furthermore, for small coupling constants, it is possible to see more pronounced radiation in the evolution.

The scalar potential was modified to turn the normal mode of the kink into a quasinormal mode. In this case, the vibrational mode of the toy model with $\epsilon=0$ was considered as the initial condition of the perturbation. Then, we evolved the perturbation numerically according to the time-dependent stability equation. When the normal mode becomes a quasinormal mode, the perturbation amplitude decays over time because it leaks energy. This causes the transition rate to also decrease over time and, consequently, the probability reaches an asymptotic value for large times.
Interestingly, the transition to the continuum is only significant for coupling constants smaller than the oscillation frequency, which is a reminiscent behavior of Fermi's golden rule for the normal mode case. These effects were observed in the numerical simulations and compared to results from first-order perturbation theory. We included the full evolution of the scalar field perturbation in the perturbation theory to have a good agreement between the two. However, we have also considered the approximation, neglecting the radiation in the scalar field because, in this case, it is possible to find the time dependence of the transition probabilities explicitly. The latter approximation gives a reasonable estimation of the asymptotic value of the transition probability.

Our work clarifies that the fermion decouples from the wobbling kink if the coupling constant is small and that this effect is attenuated if we have quasinormal instead of normal modes. This occurs because the perturbation decreases over time in the former case. Moreover, if it is possible to measure the fermion escape rate, one could, in principle, probe whether the kink is wobbling or set a bound on the wobbling parameters, such as the wobbling amplitude, quasinormal mode decay rate, and the coupling constant $g$. This could be relevant for electrons in polyacetylene \cite{su1979solitons, niemi1986fermion} and in the scenario where we live in a domain wall \cite{rubakov1983we}, to name a few examples.

\section*{Acknowledgements}

JGFC acknowledges financial support from the Brazilian agency CNPq. AM thanks financial support from the Brazilian agencies, CAPES and CNPq  Grant No. 309368/2020-0, and also Universidade Federal de Pernambuco Edital Qualis A.

\appendix

\section{Fermion Bound States of the $\phi^4$ model}
\label{ap1}

The bound and scattering states of a fermion coupled to a $\phi^4$ kink are listed in references \cite{chu2008fermions,charmchi2014complete}. The effective fermionic potentials, in this case, are given by eq.~(\ref{eq_PT}), which are P\"{o}schl-Teller potentials. In this case, the number of bound states is given by the largest integer smaller than $g$. The bound states are as follows
\begin{align}
\psi_{n,+}&=N_+^n \, \text{sech}^{g-n}(x)\, F\left(-n,2g-n+1,g-n+1;\frac{1}{2}(1-\tanh(x))\right), \nonumber\\
\psi_{n,-}&=N_-^n \, \text{sech}^{g-n}(x)\, F\left(-n+1,2g-n,g-n+1;\frac{1}{2}(1-\tanh(x))\right),
\end{align}
where $N_\pm^n$ are normalization constants obeying $N_-^0=0$ and, for $n\geq 1$, $N_-^n/N_+^n=n/E_n$. Also $E_n=\sqrt{n(2g-n)}$ is the energy of the state and $F$ is the hypergeometric function.
On the other hand, the scattering states of an incoming incident wave from the left and from the right are
\begin{align}
\psi^{\pm}_{k,L}&=N^{k,L}_{\pm}\cosh^{ik}(x)\, F\left(\frac{1}{2}-ik-\zeta_\pm,\frac{1}{2}-ik+\zeta_\pm,1-ik;\frac{1}{1+e^{2x}}\right),\nonumber\\
\psi^{\pm}_{k,R}&=N^{k,R}_{\pm}\cosh^{ik}(x)\, F\left(\frac{1}{2}-ik-\zeta_\pm,\frac{1}{2}-ik+\zeta_\pm,1-ik;\frac{1}{1+e^{-2x}}\right),
\end{align}
respectively, 
where $\zeta_{\pm}=g\pm\frac{1}{2}$, and $N^{k,L}_{\pm}$ and $N^{k,R}_{\pm}$ are normalization factors obeying $N^{k,L}_-/N^{k,L}_+=-N^{k,R}_-/N^{k,R}_+=(ik+g)/\sqrt{k^2+g^2}$. Both scattering states have energy $E_k=\sqrt{g^2+k^2}$.

\section{Numerical technique}
\label{ap2}

To solve the equations of motion numerically for the fermion field, we discretize space on the interval $-L<x<L$ with spacing $h=0.1$ and $L=100.0$. We substitute the partial derivative with respect to $x$ by a five-point stencil approximation
\begin{equation}
\frac{\partial\psi}{\partial x}=\frac{-\psi(x+2h)+8\psi(x+h)-8\psi(x-h)+\psi(x-2h)}{12h},
\end{equation}
which is fourth order accurate in $h$. At $x=-L+h$, the finite difference expression is as follows
\begin{equation}
\frac{\partial\psi}{\partial x}=\frac{\psi(-L+4h)-6\psi(-L+3h)+18\psi(-L+2h)-10\psi(-L+h)-3\psi(-L)}{12h}
\end{equation} 
where we impose $\psi(x=-L)=0$. Similar expressions are used at $x=L$. After discretizing the space, we integrate the resulting ordinary differential equation using a fifth-order Runge-Kutta method \cite{dormand1980family} implemented by the solve\_ivp method of Python library SciPy. For the scalar field, we use the analytical expression for the kink and, whenever possible, for the perturbation. This way, we ignore the higher-order coupling of the shape mode with the radiation mode \cite{manton1997kinks}. We cannot use the analytical expression for the quasinormal modes, and we must integrate the time-dependent stability eq.~(\ref{eq_time_stab}). In this case, we follow the same procedure to integrate the fermion field, i.e., the spacetime is discretized the same way, partial derivatives over $x$ are obtained using a five-point-stencil approximation, and integration over time is done using a fifth-order Runge-Kutta method.

\end{document}